\documentstyle[aps,prd,tighten,epsf,multicol]{revtex}
\begin{document}
\renewcommand{\thefootnote}{\fnsymbol{footnote}}
\title{Nonlinear electrodynamics and FRW cosmology}
\author{V. A. De Lorenci,$^{1,}$%
\footnote{Email address: lorenci@efei.br} 
R. Klippert,$^{1,}$%
\footnote{Email address: klippert@efei.br} 
M. Novello,$^{2,}$%
\footnote{Email address: novello@cbpf.br} 
and J. M. Salim$^{2,}$%
\footnote{Email address: jsalim@cbpf.br}}
\address{$^1$\,Instituto de Ci\^encias - 
Escola Federal de Engenharia de Itajub\'a\\
Av.\ BPS 1303 Pinheirinho, 37500-903 Itajub\'a, MG -- Brazil\\
%{\em(}{\tt lorenci@efei.br, klippert@efei.br}{\em)}\\
$^2$\,Centro Brasileiro de Pesquisas F\'\i sicas - CBPF\\
Rua Dr.\ Xavier Sigaud 150, 22290-180 Rio de Janeiro, RJ -- Brazil%
%\\{\em(}{\tt novello@cbpf.br, jsalim@cbpf.br}{\em)}
}
\date{Received 7 March 2001; Accepted 13 November 2001; 
Printed 15 February 2002}

\maketitle
\renewcommand{\thefootnote}{\arabic{footnote}}

\begin{abstract}
Maxwell electrodynamics, considered as a source 
of the classical Einstein field equations, 
leads to the singular isotropic Friedmann solutions. 
We show that this singular behavior does not occur for a class 
of nonlinear generalizations of the electromagnetic theory. 
A mathematical toy model is proposed for which the analytical 
nonsingular extension of FRW solutions is obtained.  
\end{abstract}

%\pacs{
{DOI:10.1103/PhysRevD.65.063501 (2002)\hfill 
PACS number(s): 98.80.Bp; 11.10.Lm}

\begin{multicols}{2}
\section{Introduction}
The standard cosmological model, 
based on Friedmann-Robertson-Walker (FRW) geometry 
with Maxwell electrodynamics as its source, 
leads to a cosmological singularity 
at a finite time in the past \cite{Kolb}.  
Such a mathematical singularity itself shows that, 
around the very beginning, 
the curvature and the energy density are arbitrarily large, 
thus being beyond the domain of applicability of the model.  
This difficulty raises also secondary problems, 
such as the horizon problem: the Universe seems to be too homogeneous 
over scales which approach its causally correlated region 
\cite{Brandenberger}.  
These secondary problems are usually solved 
by introducing geometric scalar fields 
(for a review on this approach see Ref.\ \cite{Linde} 
and references therein).  

There are many proposals of cosmological solutions 
without a primordial singularity.  
Such models are based on a variety of distinct mechanisms, 
such as cosmological constant \cite{DeSitter}, 
nonminimal couplings \cite{Gunzig}, nonlinear Lagrangians 
involving quadratic terms in the curvature \cite{Mukhanov}, 
modifications of the geometric structure of spacetime \cite{Elbaz}, 
and nonequilibrium thermodynamics \cite{Salim}, among others. 
Recently, an inhomogeneous and anisotropic nonsingular model 
for the early universe filled with a Born-Infeld-type 
nonlinear electromagnetic field was presented \cite{Breton}.  
Further investigations on regular cosmological solutions 
can be found in Ref.\ \cite{Klippert}.   

In this paper it is shown that homogeneous and isotropic 
nonsingular FRW solutions can be obtained by considering 
a toy model generalization of Maxwell electrodynamics, 
here presented as a local covariant and gauge-invariant Lagrangian 
which depends on the field invariants up to the second order, 
as a source of classical Einstein equations. 
This modification 
is expected to be relevant when the fields reach large values, 
as occurs in the primeval era of our Universe.  
Singularity theorems \cite{Hawking} are circumvented by 
the appearance of a high (but nevertheless finite) 
negative pressure in the early phase of FRW geometry. 
In the Appendix we consider the influence of other kinds of matter 
on the evolution of the universe.  
It is shown that standard matter, 
even in its ultrarelativistic state, 
is unable to modify the regularity of the obtained solution.  

Heaviside nonrationalized units are used.  
Latin indices run in the range $(1,2,3)$ 
and Greek indices run in the range $(0,1,2,3)$.  
The volumetric spatial average of an arbitrary quantity $X$ 
for a given instant of time $t$ is defined as  
\begin{equation}
\label{Ref-1}
\overline{X} \stackrel{.}{=} \lim_{V\rightarrow V_o}
\frac{1}{V}\int X\,\sqrt{-g}\,d^3\!x^i,
\end{equation}
where $V=\int\sqrt{-g}\,d^3\!x^i$, 
and $V_o$ stands for the time dependent volume of the whole space.  

\section{Einstein-Maxwell Singular Universe}
Maxwell electrodynamics usually leads to singular universe models.
In a FRW framework, this is a direct consequence of the 
singularity theorems \cite{Hawking}, 
and follows from the exam of the energy conservation law
and Raychaudhuri equation \cite{Raychaudhuri}. 
Let us set the line element 
\begin{equation}
ds^2 = c^2\,dt^2 - \frac{A^2(t)}{1+\epsilon r^2/4c^2}\,
\left[dr^2+r^2\,(d\theta^2+\sin^2\theta\,d\varphi^2)\right],
\end{equation}
where $\epsilon=-1,\,0,\,+1$ hold for 
the open, flat (or Euclidean) and closed cases, respectively.  
The 3-dimensional surface of homogeneity $t=const$ 
is orthogonal to a fundamental class of observers 
represented by a four-velocity vector field 
$v^{\mu} = c\delta^{\mu}_{o}$.  
For a perfect fluid with energy density $\rho$ and pressure $p$, 
the two above-mentioned equations assume the form
\begin{equation}
\dot{\rho} + 3(\rho + p) \frac{\dot{A}}{A} = 0,
\label{dotRho}
\end{equation}
\begin{equation}
\frac{\ddot{A}}{A} = - \,\frac{k}{6} (\rho + 3p),
\label{dotTheta}
\end{equation}
in which $k$ is the Einstein gravitational constant 
and the overdot denotes Lie derivative respective to $v$, 
that is $(1/c)\partial/\partial t$.  
Equations (\ref{dotRho}) and (\ref{dotTheta}) 
do admit a first integral 
\begin{equation}
\protect\label{constraint}
\frac{k}{3}\,\rho=\left(\frac{\dot{A}}{A}\right)^2
+\frac{\epsilon}{A^2}.
\end{equation}

Since the spatial sections of FRW geometry are isotropic, 
electromagnetic fields can generate such a universe 
only if an averaging procedure is performed\cite{Tolman}. %,Hindmarth}  
The standard way to do this is just to set 
for the electric $E_{i}$ and magnetic $H_{i}$ fields 
the following mean values: 
\begin{eqnarray}
\overline{\rule{0pt}{2ex}E_i} = 0,\qquad
%\label{meanE}\\[1ex]
%
\overline{\rule{0pt}{2ex}H_i} &=& 0,\qquad
%\label{meanH}\\[1ex]
%
\overline{\rule{0pt}{2ex}E_i\, H_j} = 0,
\label{meanEH}\\[1ex]
\overline{\rule{0pt}{2ex}E_i\,E_j} &=& -\, \frac{1}{3} E^2 \,g_{ij},
\label{meanE2}\\[1ex]
\overline{\rule{0pt}{2ex}H_i\, H_j} 
&=&  -\, \frac{1}{3} H^2 \,g_{ij}.
\label{meanH2}%\\[1ex]
\end{eqnarray}

The energy-momentum tensor associated 
with Maxwell Lagrangian is given by 
\begin{equation}
T_{\mu\nu} = F_{\mu\,\alpha}\,F^{\alpha}\mbox{}_{\nu} 
+ \frac{1}{4} \,F \,g_{\mu\nu}, 
\label{Maxwell}
\end{equation}
in which $F \stackrel{.}{=} F_{\mu\nu}\, F^{\mu\nu}=2(H^2-E^2)$.  
Using the above average values 
it follows that Eq.\ (\ref{Maxwell}) 
reduces to a perfect fluid configuration 
with energy density $\rho_\gamma$ and pressure $p_\gamma$ as 
\begin{equation}
\overline{\rule{0pt}{2ex}T_{\mu\nu}} 
= (\rho_\gamma + p_\gamma)\, v_{\mu}\, v_{\nu} 
- p_\gamma\, g_{\mu\nu},
\label{Pfluid}
\end{equation}
where
\begin{equation}
\label{RhoMaxwell}
\rho_\gamma = 3p_\gamma = \frac{1}{2}\,(E^2 + H^2).
\end{equation}
The fact that 
both the energy density and the pressure are positive definite 
for all time yields, 
using the Raychaudhuri Eq.\ (\ref{dotTheta}), 
the singular nature of FRW universes. 
Thus Einstein equations 
for the above energy-momentum configuration yield \cite{Robertson} 
\begin{equation}
\label{A(t)Maxwell}
A(t)=\sqrt{A_o^2t-\epsilon t^2},
\end{equation}
where $A_o$ is an arbitrary constant.  

\section{Nonsingular FRW universes}
\protect\label{Nonlinear}
The toy model generalization of Maxwell electromagnetic Lagrangian 
will be considered up to second order terms 
in the field invariants $F$ 
and $G \stackrel{.}{=} %F^{\star}_{\mu\nu}\,F^{\mu\nu}=
\frac{1}{2}\eta_{\alpha\beta\mu\nu}F^{\alpha\beta}F^{\mu\nu}
=-4\vec{E}\cdot\vec{H}$ as  
\begin{equation}
L = -\frac{1}{4}\,F + \alpha\,F^2 + \beta\,G^2, 
\label{Order2}
\end{equation}
where $\alpha$ and $\beta$ are arbitrary constants.   
%are to be set in order to agree with experiments.  
Maxwell electrodynamics can be formally obtained 
from Eq.\ (\ref{Order2}) by setting $\alpha=0=\beta$.  
Alternatively, it can also be dynamically obtained 
from the nonlinear theory in the limit of small fields.  
We will not consider generalizations of Eq.\ (\ref{Order2}) 
which include the term $FG$ in order to preserve parity.  
The energy-momentum tensor 
for nonlinear electromagnetic theories %\cite{Novello} 
reads 
\begin{equation}
\protect\label{Tmunu}
T_{\mu\nu}=-4\,L_F\,F_\mu\mbox{}^\alpha F_{\alpha\nu}
+ (GL_G-L)\,g_{\mu\nu},
\end{equation}
in which $L_{F}$ represents the partial derivative of the
Lagrangian with respect to the invariant $F$ and 
similarly for the invariant $G$.  
In the linear case, expression 
(\ref{Tmunu}) reduces to the usual form (\ref{Maxwell}).  

Since we are interested mainly in the analysis of the behavior 
of this system in the early universe, where matter should be 
identified with a primordial plasma %
\cite{Tajima,%Giovannini,
Campos}, we are led to limit 
our considerations to the case in which only 
the average of the squared magnetic field $H^2$ survives %
\cite{Dunne,Tajima,%Giovannini,
Joyce}.  
This is formally equivalent to put $E^2=0$ in Eq.\ (\ref{meanE2}), 
and physically means to neglect bulk viscosity terms 
in the electric conductivity of the primordial plasma.  

The homogeneous Lagrangian (\ref{Order2}) 
requires some spatial averages over large scales, 
as given by Eqs.\ (\ref{meanEH})--(\ref{meanH2}).  
If one intends to make similar calculations on smaller scales 
then either more involved non homogeneous Lagrangians should be used 
or some additional magnetohydrodynamical effect \cite{Thompson} %,Subramanian} 
should be devised in order to achieve correlation \cite{Jedamzik} 
at the desired scale.  
Since the average procedure is independent of the equations 
of the electromagnetic field we can use 
the above formulas (\ref{meanEH})--(\ref{meanH2}) 
to arrive at a counterpart of expression (\ref{Pfluid}) 
for the non-Maxwellian case.  
The average energy-momentum tensor 
is identified as a perfect fluid (\ref{Pfluid}) 
with modified expressions for the energy density $\rho_\gamma$ 
and pressure $p_\gamma$ as
\begin{eqnarray}
\rho_\gamma &=& \frac{1}{2} \, H^2 \,(1 - 8\,\alpha\,H^2),
\label{rho}\\[1ex]
\protect\label{P}
p_\gamma &=& \frac{1}{6} \,H^2 \,(1 - 40\,\alpha\,H^2).
\end{eqnarray}

Inserting expressions (\ref{rho})--(\ref{P}) 
in Eq.\ (\ref{dotRho}) yields 
\begin{equation}
H=\frac{H_o}{A^2},
\protect\label{H->A}
\end{equation}
where $H_o$ is a constant.  
%Equation (\ref{H->A}) 
%This relation holds as well in the Maxwell case.  
With this result, 
a similar procedure applied to Eq.\ (\ref{constraint}) leads to 
\begin{equation}
\label{eqA2}
\dot{A}^2=\frac{kH_o^2}{6\,A^2}
\left(1-\frac{8\alpha H_o^2}{A^4}\right)-\epsilon.
\end{equation}
As far as 
the right-hand side of Eq.\ (\ref{eqA2}) must not be negative 
it follows that, regardless of the value of $\epsilon$, 
for $\alpha>0$ the scale factor $A(t)$ 
cannot be arbitrarily small.  
The solution of Eq.\ (\ref{eqA2}) is implicitly given as 
\begin{equation}
\label{solution}
c\,t=\pm\int_{A_o}^{A(t)}\frac{\textstyle dz}
{\sqrt{\textstyle\frac{\textstyle kH_o^2}{\textstyle6z^2}
-\frac{\textstyle8\alpha kH_o^4}{\textstyle6z^6}-\epsilon}},
\end{equation}
where $A(0)=A_o$.  
The linear case (\ref{A(t)Maxwell}) 
can be achieved from Eq.\ (\ref{solution}) by setting $\alpha=0$.

\begin{figure}[htb]
\leavevmode
\centering
\epsfysize=50ex
\epsfbox{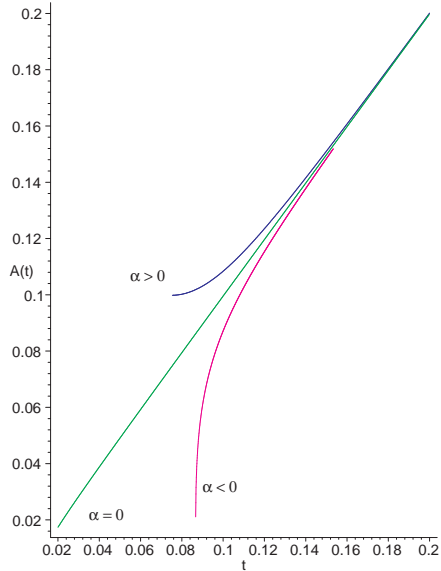}
%\protect\vspace{-7\baselineskip}
\caption{Plots from Eq.~(\protect\ref{solution}).  
We set $A(1)=1$, $kH_o^2=12$, 
and $\alpha H_o^2=(0;\,\pm 1.25\times10^{-4})$ as illustrative values.}
\label{Fig3}
\end{figure}

\end{multicols}
A closed form of Eq. (\ref{solution}) 
for $\epsilon=\pm1$ can be derived as
\begin{equation} 
\label{solution+-} 
c\,t=\pm\left[\frac{(x_1-x_3){\cal E}
\left(\arcsin\sqrt{\frac{\textstyle z-x_1}{\textstyle x_2-x_1}},
\sqrt{\frac{\textstyle x_1-x_2}{\textstyle x_1-x_3}}\right) 
+x_3{\cal F}
\left(\arcsin\sqrt{\frac{\textstyle z-x_1}{\textstyle x_2-x_1}},
\sqrt{\frac{\textstyle x_1-x_2}{\textstyle x_1-x_3}}\right)}
{\sqrt{x_3-x_1}}\right]_{z=A_o^2}^{z=A^2(t)}, 
\label{referee}
\end{equation} 
where $x_1,\,x_2,\,x_3$ are the three roots of the equation  
$8\alpha kH_o^4-kH_o^2x+3\epsilon x^3=0$, and
%${\cal F}$ and ${\cal E}$ 
\begin{displaymath}
{\cal F}(x,\,\kappa)\doteq
\int\limits_{\textstyle0}^{\textstyle\sin\,x}
\frac{\textstyle1}{\textstyle\sqrt{(1-z^2)(1-\kappa^2z^2)}}\,dz,\qquad 
{\cal E}(x,\,\kappa)\doteq
\int\limits_{\textstyle0}^{\textstyle\sin\,x}
\frac{\textstyle\sqrt{1-\kappa^2z^2}}{\textstyle\sqrt{1-z^2}}\,dz
\end{displaymath}
are the elliptic functions  
of the first and of second kind, respectively 
(see expressions 8.111.2 and 8.111.3 in Ref.\ \cite{Gradshteyn}).   
The behavior of $A(t)$ for $\epsilon=\pm1$
is displayed in the Fig.\ \ref{Fig3}.
\begin{multicols}{2}
For the Euclidean section, by suitably choosing the origin of time, 
expression (\ref{solution}) can be solved as 
\begin{equation}
\label{A(t)}
A^2 = H_{o} \,\sqrt{\frac{2}{3} \,(k\,c^2\,t^2 +12\,\alpha)}.
\end{equation}
>From Eq.\ (\ref{H->A}), 
the average strength of the magnetic field $H$ evolves in time as
\begin{equation}
\protect\label{H(t)}
H^2 = \frac{3}{2}\,\frac{1}{{k\,c^2\,t^2} + 12\,\alpha}.
\end{equation} 

\begin{figure}[htb]
\leavevmode
\centering
\epsfysize=50ex
\mbox{\hspace*{-1cm}
\epsfbox{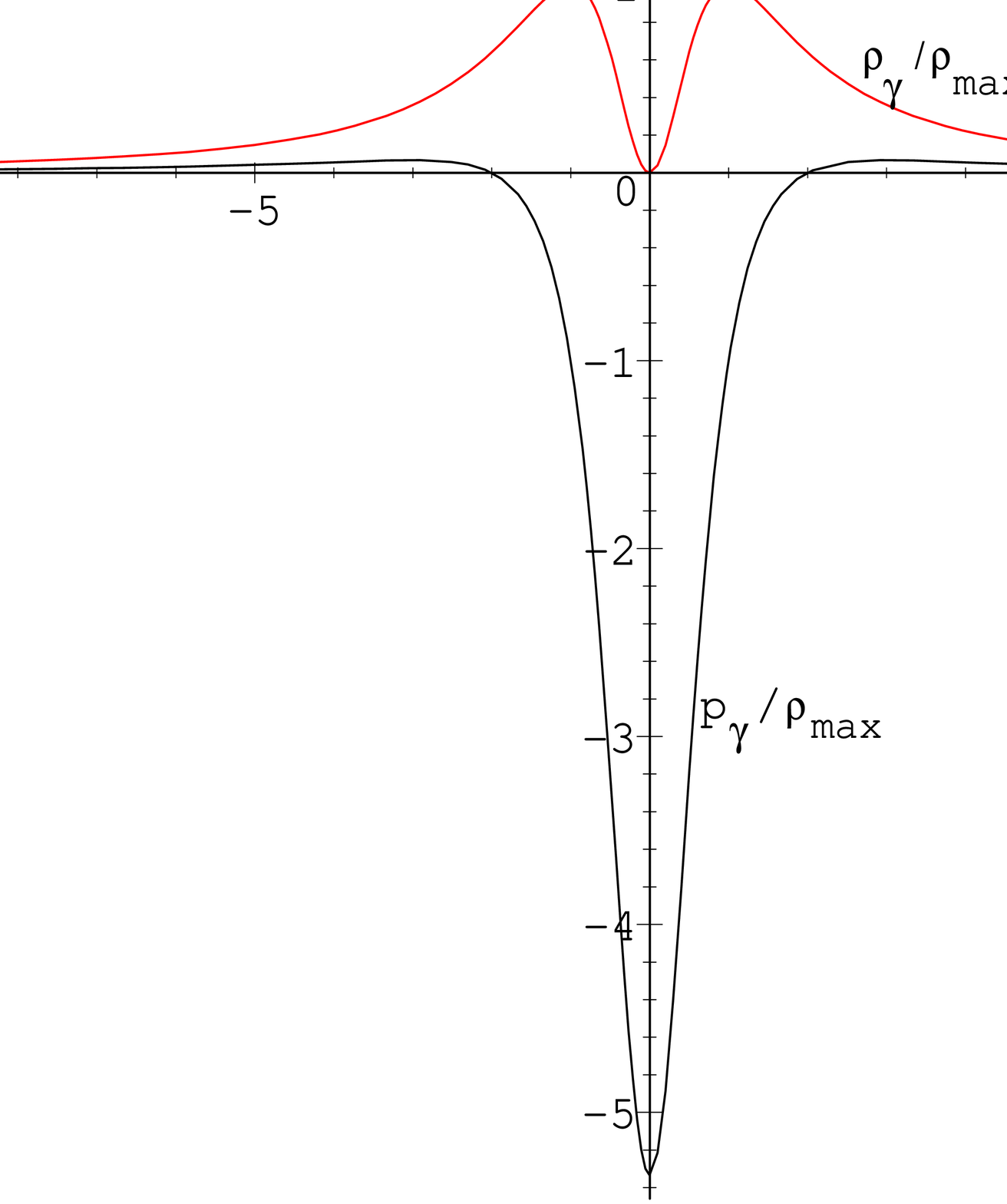}
\hspace{3cm}
\epsfysize=40ex
\epsfbox{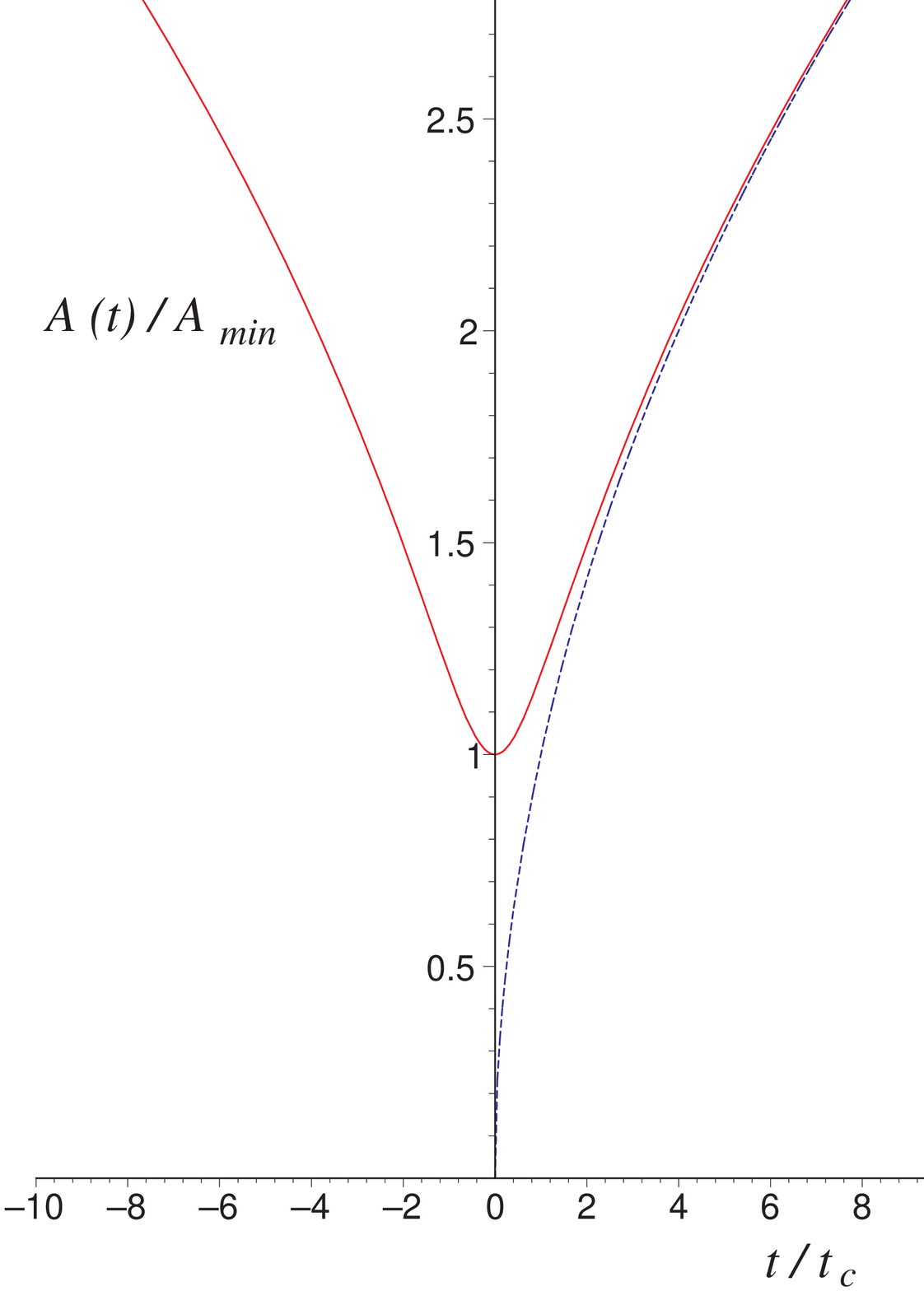}}
%\protect\vspace{-7\baselineskip}
\caption{On the left panel: 
time dependence of the electromagnetic energy density $\rho_\gamma$ 
and pressure $p_\gamma$.  %}
$\rho_{max}=1/64\alpha$ %is given from (\protect\ref{maxRho}) 
and $t_c$ is given by Eq.\ (\protect\ref{tc}).  %}
On the right panel: 
nonsingular behavior of the scale factor $A(t)$.  
$A_{min}$ and $t_c$ are given from Eqs.\ 
(\protect\ref{Amin}),(\protect\ref{tc}). 
The corresponding classical expression (\protect\ref{A(t)Maxwell}) 
is shown (dashed line) for comparison, with $A_o=A_{min}$.  }
\label{FigP(t)}
\end{figure}

Expression (\ref{A(t)}) is singular for $\alpha<0$, 
as there exist a time $t=\sqrt{-12\alpha/k\,c^2}$ 
for which $A(t)$ is arbitrarily small.  
Otherwise, for $\alpha>0$ we recognize that 
at $t = 0$ the radius of the universe attains a minimum value 
%\footnote{For alternative models with bouncing FRW solutions 
%see references \cite{Balbinot}.}%,Bekenstein,Murphy,Randibar,
%Mannheim,Alvarenga,Dirar,Gasperini}} 
$A_{min}$, which is given from 
\begin{equation}
\label{Amin}
A^2_{min} = H_{o} \, \sqrt{8\,\alpha}.
\end{equation}
Therefore, the actual value of $A_{min}$ depends on $H_o$, 
which turns out to be the unique free parameter 
of the present model.  
The energy density $\rho_\gamma$ given by Eq.\ (\ref{rho}) 
reaches its maximum value $\rho_{max}=1/64\alpha$ 
%\begin{equation}
%\protect\label{maxRho}
%\rho_{max} = \frac{1}{16 \,\mu}
%\end{equation}
at the instant $t=t_c$, where
\begin{equation}
\label{tc}
t_{c} = \frac{1}{c} \,\sqrt{\frac{12\,\alpha}{k}}.
\end{equation}
For smaller values of $t$ the energy density decreases, 
vanishing at $t = 0$, 
while the pressure becomes negative. 
%As remarked before, 
Only for times 
$t\,\raisebox{-0.5ex}{$\stackrel{<}{\scriptsize\sim}$}
\,10\sqrt{\alpha/kc^{2}}$ 
the nonlinear effects are relevant 
for cosmological solution of the normalized scale factor, 
as shown in Fig.\ \ref{FigP(t)}.  
Indeed, solution (\ref{A(t)}) 
fits the standard expression (\ref{A(t)Maxwell}) 
of the Maxwell case at the limit of large times.

\end{multicols}
%%
%\begin{figure}[htb]
%\leavevmode
%\centering
%%\epsfxsize=80ex
%\epsfysize=50ex
%\epsfbox{medfig1b.eps}
%%\protect\vspace{-7\baselineskip}
%\caption{nonsingular behavior of the scale factor $A(t)$.  
%$A_{min}$ is given from (\protect\ref{Amin}) 
%and $t_c$ from (\protect\ref{tc}). 
%The classical expression (\protect\ref{A(t)Maxwell}) 
%is shown (dashed line) for comparison, 
%with $A_o=A_{min}$.}
%\label{FigA(t)}
%\end{figure}
\begin{multicols}{2}

The energy-momentum tensor (\ref{Tmunu}) 
is not trace-free for $\alpha\neq0$.  
Thus, the equation of state 
\begin{math}
p_\gamma = p_\gamma(\rho_\gamma)
\end{math}
is no longer given by the Maxwellian value; 
it has instead a quintessential-like term \cite{Caldwell} 
which is proportional to the constant $\alpha$.  That is
\begin{equation}
\label{newstate}
p_\gamma = \frac{1}{3} \,\rho_\gamma - \frac{16}{3}\,\alpha \,H^4.
\end{equation}
Equation (\ref{newstate}) can also be written in the form 
\begin{eqnarray}
p_\gamma &=& \frac{1}{3}\,\rho_\gamma-\frac{1}{24\alpha}
\left\{(1-32\alpha\rho_\gamma)\right.\nonumber\\&&\left.
+[1-2\Theta(t-t_c)]\sqrt{1-64\alpha\rho_\gamma}\right\},
\label{Ref-2}
\end{eqnarray}
where $\Theta(z)$ it the Heaviside step function.  
The right-hand side of Eq.\ (\ref{Ref-2}) 
behaves as $(1-64\alpha\rho_\gamma)\rho_\gamma/3$ 
for $t>t_c$ in the Maxwell limit $\alpha\rho_\gamma\ll 1$.

The maximum of the temperature corresponding to $t=t_c$ is given by 
\begin{equation}
\protect\label{Tmax}
T_{max}=\left(\frac{c}{24\alpha\sigma}\right)^{1/4},
\end{equation}
where $\sigma$ is the Stefan-Boltzmann constant.  

\section{Conclusions}
The consequences of the minimal coupling of gravity 
with second order nonlinear electrodynamics (\ref{Order2}) 
were examined.  
>From the cosmological point of view, 
the proposed modification is relevant only 
in the primeval era of the universe.  
Indeed, the class of theories $\alpha>0$ 
leads to nonsingular solutions for which 
the scale factor $A(t)$ attains a minimum value.  
The regularity of the cosmological solution (\ref{A(t)}) 
is to be attributed to the fact that, for $t<\sqrt{24\alpha/kc^2}$, 
the quantity $\rho+3p$ becomes negative.

\section*{Acknowledgments}
This work was partially supported by 
{\em Conselho Nacional de Desenvolvimento Cient\'{\i}fico 
e Tecnol\'ogico} (CNPq) and 
{\em Funda\c{c}\~ao de Amparo \`a Pesquisa 
do Estado de Minas Gerais} (FAPEMIG)
of Brazil.

\appendix
\section{Ultrarelativistic matter contribution}
Beside photons there are plenty of other particles, 
and physics of the early universe deals with various sort of matter. 
In the standard framework they are treated in terms of a fluid 
% and pressure $p_\nu$
with energy density $\rho_{\nu}$, which satisfies 
an ultrarelativistic equation of state $p_{\nu} = \rho_{\nu}/3$.  
Adding the contribution of this kind of matter to 
the average energy-momentum tensor 
%$\overline{\rule{0pt}{2ex}T_{\mu\nu}}$ 
of the photons 
%and neglecting the interacting term 
it follows that 
\begin{math}
%\label{otherRho}
\rho_{\nu} = K A^{-\,4},
\end{math} 
where $K$ is an arbitrary positive constant.  
This result allows us to treat such extra matter as nothing but 
a reparametrization of the constants $H_{o}$ and $\alpha$ 
as $\hat{H}_{o}^2 = H_{o}^2 + 2K$ and 
$\hat{\alpha}=\alpha{H_{o}^{4}}/{(H_o^2+2K)^2}$.
The net effect of this 
is just to diminish the value of $A_{min}$ as 
\begin{equation}
\label{newA}
\hat{A}\mbox{}_{min}=
\left(\frac{H_o^2}{H_o^2+2K}\right)^{1/4}A_{min}.
\end{equation}
Therefore, it turns out that the phenomenon of 
reversing the sign of the expansion factor $3\dot{A}/A$ 
due to the high negative pressure of the photons 
is not essentially modified by the ultrarelativistic gas.  
Only an exotic fluid possessing energy density 
$\rho_{exotic} \sim A^n$ with $n \le -\,8$ 
could be able to modify the above result.  
However, this situation seems to be a very unrealistic case.

\end{multicols}

\begin{thebibliography}{28}
\bibitem{Kolb}
E. W. Kolb and M. S. Turner, {\em The Early Universe} 
(Addison-Wesley, Redwood City, CA, 1990).
\bibitem{Brandenberger}
R. Brandenberger, in {\em Proceedings of the 
VIII Brazilian School of Cosmology and Gravitation}, 
edited by M. Novello (Editions Fronti\'eres, Singapore, 1996). 
\bibitem{Linde}
L. Kofman, A. Linde, and A. A. Starobinsky
{\em Phys.\ Rev.\ D} {\bf 56}, 3258 (1997).
\bibitem{DeSitter} 
W. de Sitter, Proc.\ K. Ned.\ Akad.\ Wet.\ {\bf 19}, 1217 (1917).
\bibitem{Gunzig}
M. Novello and J. M. Salim, {\em Phys.\ Rev. D} {\bf 20}, 377 (1979);
A. Saa, E. Gunzig, L. Brenig, V. Faraoni, 
T. M. Rocha Filho, and A. Figueiredo, 
{\tt gr-qc/0012105} (2000). 
\bibitem{Mukhanov}
V. Mukhanov and R. Brandenberger, 
{\em Phys.\ Rev.\ Lett.} {\bf 68}, 1969 (1992).
%\bibitem{Sornborger} 
See also 
R. Brandenberger, V. Mukhanov, and A. Sornborger, 
{\em Phys.\ Rev.\ D} {\bf 48}, 1629 (1993); 
R. Moessner and M. Trodden, 
{\em ibid} {\bf 51}, 2801 (1995).
\bibitem{Elbaz}
M. Novello, L. A. R. Oliveira, J. M. Salim and E. Elbaz, 
{\em Int.\ J. Mod.\ Phys.\ A} {\bf 1}, 641 (1993).
\bibitem{Salim}
G. L. Murphy, {\em Phys.\ Rev.\ D} {\bf 8}, 4231 (1973); 
J. M. Salim and H. P. de Olivera, 
{\em Acta Phys.\ Pol.\ B} {\bf 19}, 649 (1988).
\bibitem{Breton}R. Garcia-Salcedo and N. Breton, 
{\em Int.\ J. Mod.\ Phys.\ A} 
{\bf 15} (27), 4341 (2000).
\bibitem{Klippert}
R. Klippert, V. A. De Lorenci, M. Novello, and J. M. Salim, 
{\em Phys.\ Lett.\ B} {\bf 472}, 27 (2000);
%\bibitem{Veneziano}
G. Veneziano, {\tt hep-th/0002094} 2000.
\bibitem{Hawking}
S. W. Hawking and G. F. R. Ellis, 
{\em The Large Scale Structure of Spacetime} 
(Cambridge University Press, Cambridge, England, 1973); 
%\bibitem{Wald}
R. M. Wald, {\em General Relativity} 
(Univ.\ Chicago Press, Chicago, 1984).
\bibitem{Raychaudhuri}
M. Novello, in {\em Proceedings of the 
II Brazilian School of Cosmology and Gravitation}, 
edited by M. Novello (J. Sansom \& Cia., Rio de Janeiro, 1980) 
(in Portuguese).
\bibitem{Tolman} 
R. C. Tolman and P. Ehrenfest, 
{\em Phys.\ Rev.\ } {\bf 36}, 1791 (1930);
%\bibitem{Hindmarth}
M. Hindmarsh and A. Everett, 
{\em Phys.\ Rev.\ D} {\bf 58}, 103505 (1998).%;
%R. Klippert, V. A. De Lorenci, M. Novello, and J. M. Salim, 
%{\em Phys.\ Lett.\ B} {\bf 472}, 27 (2000).
\bibitem{Robertson}
H. P. Robertson, {\em Rev.\ Mod.\ Phys.} {\bf 5}, 62 (1933);
%\bibitem{Edwards}
D. Edwards, {\em Astrophys.\ Space Sci.} {\bf 24}, 563 (1973);
%\bibitem{Coqueraux}
R. Coqueraux and A. Grossmann, {\em Ann.\ Phys.} 
(N.Y.) {\bf 143}, 296 (1982);
%\bibitem{Dabrowski}
M. Dabrowski and J. Stelmach, {\em Astron.\ J.} {\bf 92}, 1272 (1986).
\bibitem{Tajima}
T. Tajima, S. Cable, K. Shibata, and R. M. Kulsrud, 
{\em Astrophys.\ J.} {\bf 390}, 309 (1992);
%\bibitem{Giovannini}
M. Giovannini and M. Shaposhnikov, 
{\em Phys.\ Rev.\ D} {\bf 57}, 2186 (1998).
\bibitem{Campos}
A. Campos and B. L. Hu, {\em Phys.\ Rev.\ D} {\bf 58}, 125021 (1998).
\bibitem{Dunne}
G. Dunne and T. Hall, {\em Phys.\ Rev.\ D} {\bf 58}, 105022 (1998); 
G. Dunne, {\em Int.\ J. Mod.\ Phys.\ A} {\bf 12}, 1143 (1997).
%\bibitem{Novello}
%M. Novello, J. M. Salim, V. A. De Lorenci, and R. Klippert, 
%{\em Phys.\ Rev.\ D} {\bf 61}, 045001 (2000); 
%V. A. De Lorenci, R. Klippert, M. Novello, and J. M. Salim, 
%{\em Phys.\ Lett.\ B} {\bf 482}, 134 (2000).
\bibitem{Joyce}
M. Joyce and M. Shaposhnikov, 
{\em Phys.\ Rev.\ Lett.\ } {\bf 79}, 1193 (1997).
\bibitem{Thompson}
C. Thompson and O. Blaes, 
{\em Phys.\ Rev.\ D} {\bf 57}, 3219 (1998);
%\bibitem{Subramanian}
K. Subramanian and J. D. Barrow, 
{\em ibid} {\bf 58}, 883502 (1998). 
\bibitem{Jedamzik}
K. Jedamzik, V. Jatalini\'c, and A. V. Olinto, 
{\em Phys.\ Rev.\ D} {\bf 57}, 3264 (1998).
\bibitem{Gradshteyn}I. S. Gradshteyn and I. M. Ryzhik, 
{\em Table of Integrals, Series, and Products}, 
(Academic, London, 1965).
\bibitem{Caldwell}
R. R. Caldwell, R. Dare, and P. J. Steinhardt, 
{\em Phys.\ Rev.\ Lett.\ } {\bf 80}, 1582 (1998).
\end{thebibliography}
\end{document}